\documentclass[runningheads]{llncs}
\usepackage{graphicx}
\usepackage{subfig}
\begin{document}

\title{HIVE-4-MAT: Advancing the Ontology Infrastructure for Materials Science\thanks{Supported by NSF Office of Advanced Cyberinfrastructure (OAC): \#1940239.}}
%
%
\author{Jane Greenberg\orcidID{0000-0001-7819-5360} \and
Xintong Zhao\orcidID{0000-0001-8401-356X} \and
Joseph Adair\orcidID{0000-0001-5646-9041} \and
Joan Boone\orcidID{0000-0001-5646-9041} \and
Xiaohua Tony Hu\orcidID{0000-0002-4777-3022}}
\authorrunning{Greenberg, et al.}
%
\institute{Metadata Research Center, Drexel University, Philadelphia PA 19104, USA \\
\email{\{jg3243,xz485,jda57,jpb357,xh29\}@drexel.edu}}
\maketitle              
\begin{abstract}
This paper introduces Helping Interdisciplinary Vocabulary Engineering for Materials Science (HIVE-4-MAT), an automatic linked data ontology application. The paper provides contextual background for materials science, shared ontology infrastructures, and knowledge extraction applications. HIVE-4-MAT’s three key features are reviewed: 1) Vocabulary browsing, 2) Term search and selection, and 3)Knowledge Extraction/Indexing, as well as the basics of named entity recognition (NER). The discussion elaborates on the importance of ontology infrastructures and steps taken to enhance knowledge extraction. The conclusion highlights next steps surveying the ontology landscape, including NER work as a step toward relation extraction (RE), and support for better ontologies.

\end{abstract}
\section{Introduction}
A major challenge in materials science research today is that the artifactual embodiment is primarily textual, even if it is in digital form. Researchers analyze materials through experiments and record their findings in textual documents such as academic literature and patents. The most common way to extract knowledge from these artifacts is to read all the relevant documents, and manually extract knowledge. However, reading is time-consuming, and it is generally unfeasible to read and mentally synthesize all the relevant knowledge\cite{tshitoyan2019unsupervised,Weston2019}. Hence, effectively extracting knowledge and data becomes a problem. One way to address this challenge is through knowledge extraction using domain-specific ontologies \cite{haendel2018classification}. Unfortunately, materials science work in this area is currently hindered by limited access to and use of relevant ontologies. This situation underscores the need to improve the state of ontology access and use for materials science research, which is the key goal of the work presented here.

This paper introduces Helping Interdisciplinary Vocabulary Engineering for Materials Science (HIVE-4-MAT), an automatic linked data ontology application. The contextual background covers materials science, shared ontology infrastructures, and knowledge extraction applications. HIVE-4-MAT’s basic features are reviewed, followed by a brief discussion and conclusion identifying next steps. 

\section{Background}
\subsection{Materials Science}

Materials science is an interdisciplinary field that draws upon chemistry, physics, engineering and interconnected disciplines. The broad aim is to advance the application of materials for scientific and technical endeavors. Accordingly, materials science researchers seek to discover new materials or alter existing ones; with the overall aim of offering more robust, less costly, and/or less environmentally harmful materials. 

Materials science researchers primarily target solid matter, which retains its shape and character compared to liquid or gas. There are four key classes of solid materials: metals, polymers, ceramics, and composites. Researchers essentially process (mix, melt, etc.) elements in a controlled way, and measure performance by examining a set of properties. Table 1 provides two high-level examples of materials classes, types, processes, and properties.

\begin{table}[]
\begin{tabular}{|l|l|l|}
\hline
\textbf{\begin{tabular}[c]{@{}l@{}}MATERIAL \\ CLASS \& TYPE\end{tabular}}                                                        & \textbf{MANUFACTURING PROCESS}                                                                                                                       & \textbf{\begin{tabular}[c]{@{}l@{}}PROPERTIES \\ (examples)\end{tabular}}                                                                        \\ \hline
\begin{tabular}[c]{@{}l@{}}Class: \\ Polymer\\ \\ Type: \\ Polyethylene\cite{rogers2015everything}\end{tabular} & \begin{tabular}[c]{@{}l@{}}Polymerization (distillation of \\ ethane into fractions, some of \\ which are combined with catalysts)\end{tabular}      & \begin{tabular}[c]{@{}l@{}}Melt temperature\\ Tensile strength\\ Flexurile strength \\ (or bend strength)\\ Shrink Rate\end{tabular}             \\ \hline
\begin{tabular}[c]{@{}l@{}}Class: Metal\\ Type: Steel\end{tabular}                                                                & \begin{tabular}[c]{@{}l@{}}Iron ore is heated and forged in \\ blast furnaces, where the impurities \\ are altered and carbon is added.\end{tabular} & \begin{tabular}[c]{@{}l@{}}Yield strength\\ Tensile strength\\ Thermal conductivity\\ Resistance to \\ wear/corrosion\\ Formability\end{tabular} \\ \hline
\end{tabular}
\vspace{2mm}
\caption{Examples of Materials classes and types, processes, and properties}
\label{table1}
\end{table}

The terms in Table \ref{table1} have multiple levels (sub-types or classes) and variants. For example, there is stainless steel and surgical steel. Moreover, the universe of properties, which is large, extends even further when considering nano and kinetic materials. This table illustrates the language, hence the ontological underpinnings, of materials science, which is invaluable for knowledge extraction. Unfortunately, the availability of computationally ready ontologies applicable to materials science is severely limited, particularly compared to biomedicine and biology. 

\subsection{Ontologies: Shared Infrastructure and Knowledge Extraction Applications}

Ontologies have provided a philosophical foundation and motivation for scientific inquiry since ancient times\cite{jg2015}. Today, computationally ready ontologies conforming to linked data standards\cite{bizer2009emerging} offer a new potential for data driven discovery\cite{eisenberg2019uncovering}. Here, the biomedical and biology communities have taken the lead in developing a shared infrastructure, through developments such as the National Center for Biological Ontologies (NCBO) Bioportal\cite{whetzel2011bioportal,NCBO} and the OBO foundry\cite{smith2007obo,OBO}. Another effort is the FAIRsharing portal\cite{sansone2019fairsharing,FAIRsharing}, providing access to a myraid of standards, databases, and other resources\cite{wilkinson2016fair}. 

Shared ontology infrastructures help standardize language and support data interoperability across communities. Additionally, the ontological resources can aid knowledge extraction and discovery. Among one of the best known applications in this area is Aronson's \cite{aronson2001effective} $MetaMap$, introduced in 2001. This application extracts key information from textual documents, and maps the indexing to the metathesaurus ontology. The $MetaMap$ application is widely-used for extraction of biomedical information. The HIVE application\cite{greenberg2011hive}, developed by the Metadata Research Center, Drexel University, also supports knowledge extraction in a same way, although results are limited by the depth of the ontologies applied. For example, biomedicine ontologies, which often have a rich and deep network of terms, will produce better results compared to more simplistic ontologies targeting materials science\cite{zhao2020scholarly,Zhang2015}.

Overall, existing ontology infrastructure and knowledge extraction approaches are applicable to materials science. In fact, biology and biomedical ontologies are useful for materials science research, and researchers have been inspired by these developments to develop materials science ontologies\cite{anikin2019ontology,Cheung2009,Greenberg2015}. Related are nascent efforts developing shared metadata and ontology infrastructures for materials science. Examples include the NIST Materials Registry\cite{NIST} and the Industrial Ontology Foundry\cite{IOF}. These developments and the potential to leverage ontologies for materials science knowledge extraction motivate our work to advance HIVE-4-MAT. They have also had a direct impact on exploring the use of NER to assist in the development richer ontologies for materials science \cite{zhao2020scholarly}.

\subsection{Named Entity Recognition}

The expanse and depth of materials science ontologies is drastically limited, pointing to a need for richer ontologies; however, ontology development via manual processes is a costly undertaking. One way to address this challenge is to through relation extraction and using computational approaches to develop ontologies. To this end, named entity recognition (NER) can serve as an invaluable first step, as explained here.

The goal of Named Entity Recognition (NER) is to recognize key information that are related to predefined semantic types from input textual documents\cite{li2020survey}. As an important component of information extraction (IE), it is widely applied in tasks such as information retrieval, text summarization, question answering and knowledge extraction. 

The semantic types can vary depending on specific task types. For example, when extracting general information, the predefined semantic types can be location, person, or organization. NER approaches have been also proven effective to biomedical information extraction; an example from SemEval2013 task 9\cite{segura-bedmar-etal-2013-semeval} about NER for drug-drug interaction is shown in Figure \ref{ner_example} below.

\begin{figure}
\centering
\includegraphics[width=0.9\textwidth]{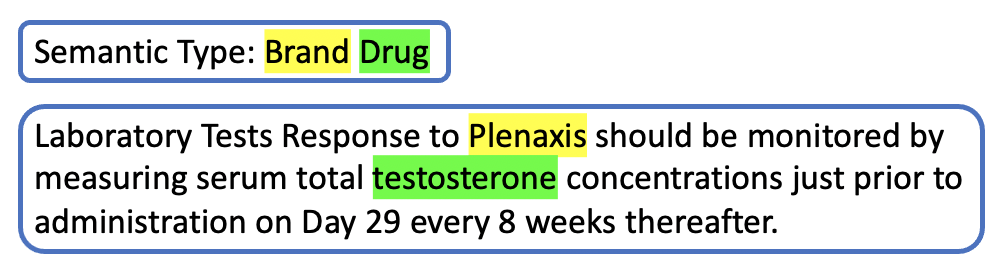}
\caption{An NER Example from a SemEval Task} 
\label{ner_example}
\end{figure}

As shown in the Figure 1, the NER pharmaceutical model receives the textual input (e.g. sentences), and returns whether there are important information entities that belong to any predefined labels, such as brand name and drug name.

A similar undertaking has been pursued by Weston et al.\cite{Weston2019}, with their NER model designed for \textbf{inorganic materials} information extraction. Their model includes seven entity labels and testing has resulted in an overall f1-score of 0.87\cite{Weston2019}. This work has inspired the HIVE team to use NER, as a step toward relation extraction, and the development of richer ontologies for materials science.

\section{Purpose and Goals}

Goals of this paper are to:

\begin{enumerate}
\item Introduce HIVE
\item Demonstrate HIVE's three key features Vocabulary browsing, term search and selection, and knowledge extraction/indexing
\item Provide an example of our NER work, as a foundation for relation extraction.
\end{enumerate}

\section{HIVE-4-MAT: Prototype Development and Features}

\begin{figure}
\centering
\includegraphics[width=0.9\textwidth]{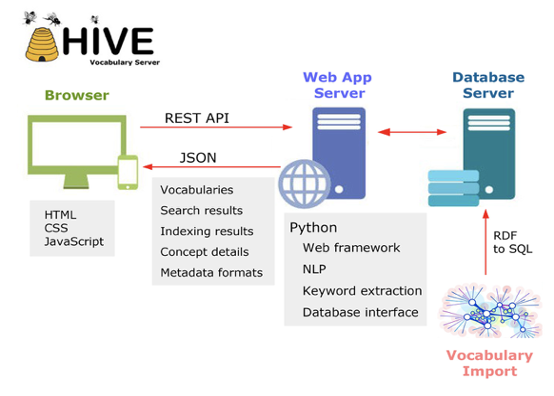}
\caption{Overview of HIVE Structure} \label{hive}
\end{figure}

Hive is a linked data automatic metadata generator tool developed initially as a demonstration for the Dryad repository\cite{greenberg2011hive,white2012hive}, and incorporated into the DataNet Federation Consortium’s iRODS system\cite{conway2013advancing}. Ontologies encoded in the Simple Knowledge Organization System (SKOS) format are shared through a HIVE-server. Currently, HIVE 2.0 uses Rapid Automatic Keyword Extraction (RAKE), an unsupervised algorithm that processes and parses text into a set of candidate keywords based on co-occurrence\cite{rose2010automatic}. Once the list of candidate keywords is selected from the SKOS encoded ontologies, the HIVE system matches candidate keywords to terms in the selected ontologies. Figure \ref{hive} provides an overview of the HIVE model.

HIVE-4-MAT builds on the HIVE foundation, and available ontologies have been selected for either broad or targeted applicability to materials science. The prototype includes the following ten ontologies: 1)Bio-Assay Ontology (BioAssay), 2) Chemical Information Ontology (CHEMINF), 3) Chemical Process Ontology (prochemical), (4) Library of Congress Subject Headings (LCSH), 5) Metals Ontology, 6) National Cancer Institute Thesaurus (NCIT), 7) Physico-Chemical Institute and Properties (FIX), 8) Physico-chemical process (REX), 9) Smart Appliances REFerence Ontology (SAREF), and 10) US Geological Survey (USGS).

\begin{figure}[h]
\centering
\includegraphics[width=0.9\textwidth]{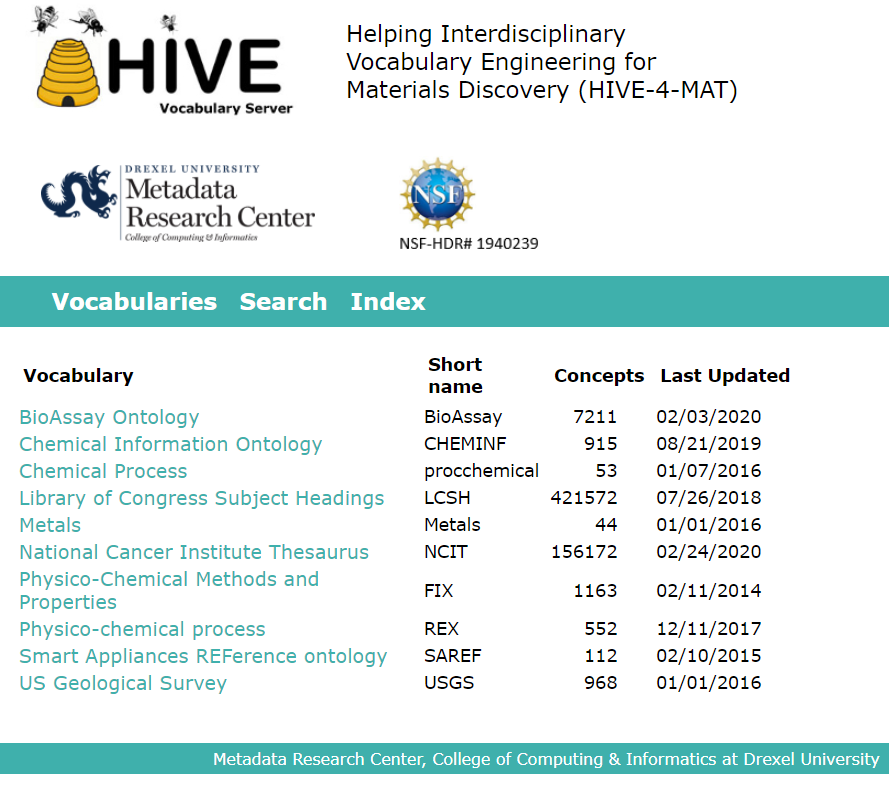}
\caption{Lists of Vocabularies/Ontologies} 
\label{fig1}
\end{figure}

\begin{figure}[]
\centering
\includegraphics[width=0.9\textwidth]{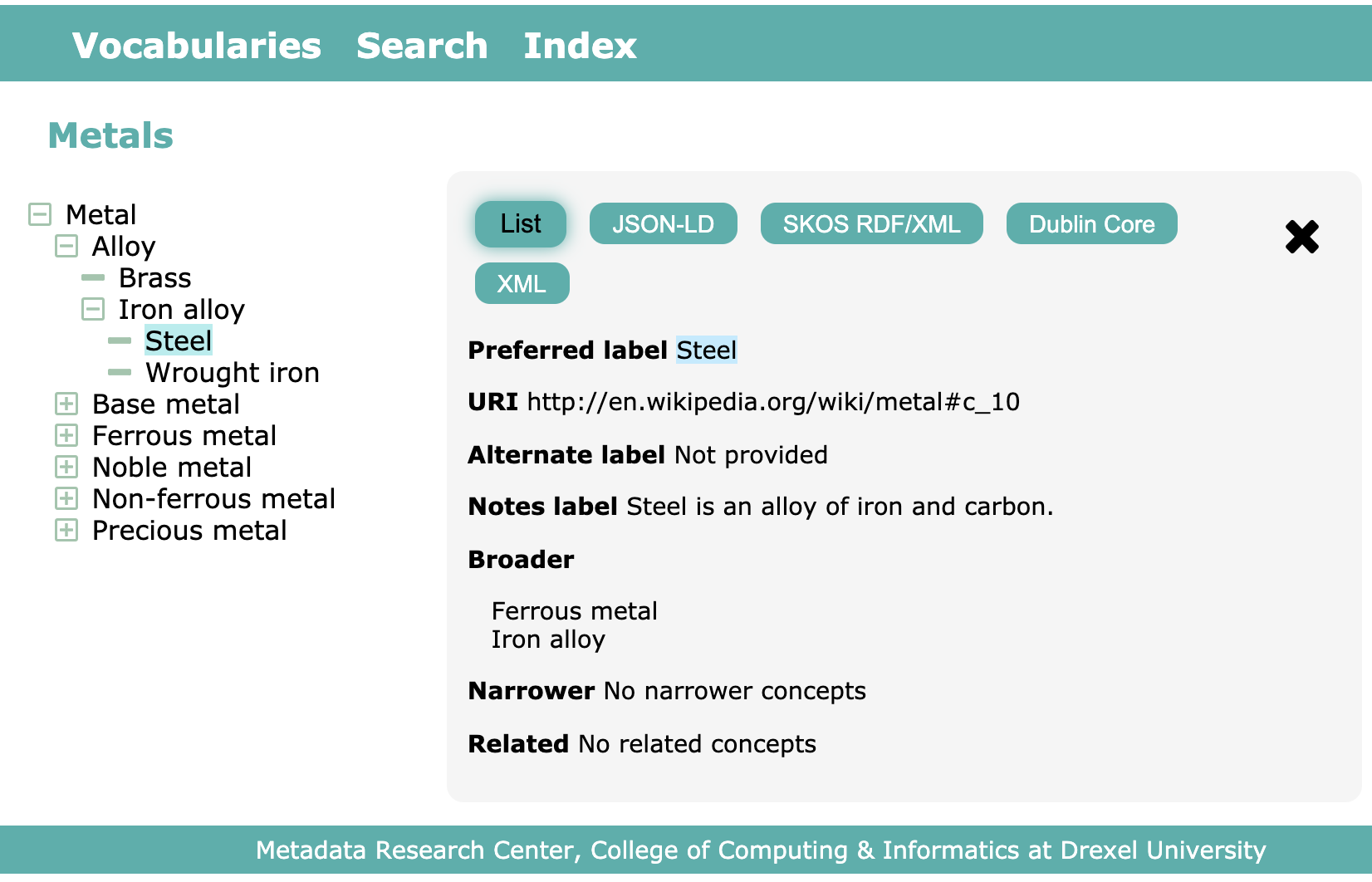}
\caption{Vocabularies/Ontologies Structure} 
\label{fig2}
\end{figure}

Currently, HIVE-4-MAT has three main features: 
\begin{itemize}
  \item[$\bullet$] Vocabulary browsing (Figure \ref{fig1} and Figure \ref{fig2})
  \item[$\bullet$] Term search and selection  (Figure \ref{fig3})
  \item[$\bullet$] Knowledge Extraction/Indexing (Figure \ref{fig4})
\end{itemize}

The vocabulary browsing feature allows a user to view and explore the ontologies registered in HIVE-4-MAT. Figure \ref{fig1} presents the full list of currently available ontologies, and Figure \ref{fig2} provides an example navigating through the hierarchy of the Metals ontology. The left-hand column (Figure \ref{fig2}) displays the hierarchical levels of this ontology; the definition, and the right-hand side displays the alternative name, broader concepts and narrow concepts.

\subsection{Mapping Input Text to Ontologies}

\begin{figure}
\centering
\includegraphics[width=0.9\textwidth]{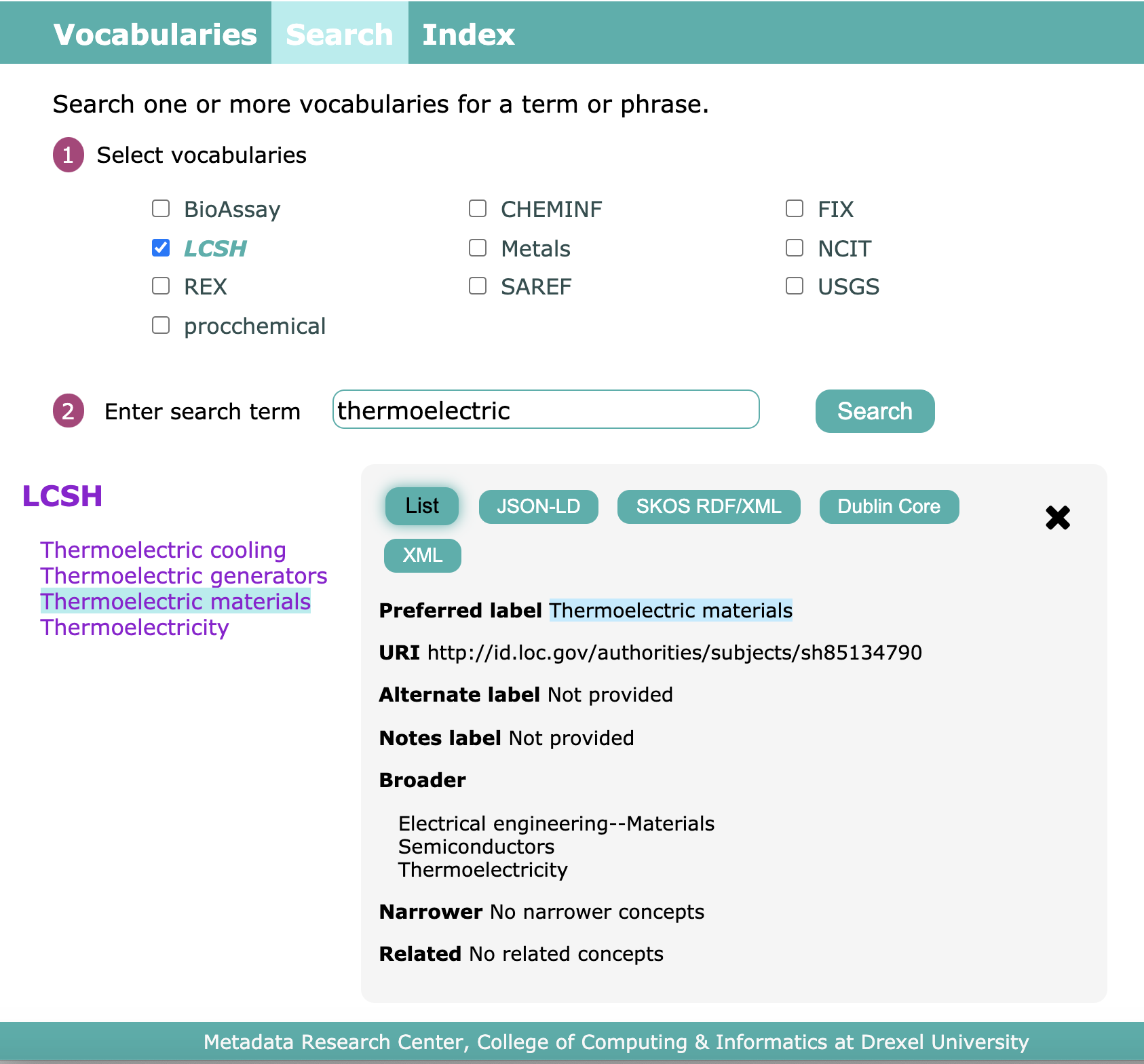}
\caption{Term Search}
\label{fig3}
\end{figure}

The term search and selection feature in Figure \ref{fig3} allows a user to select a set of ontologies and enter a search term. In this example, eight of the 10 ontologies are selected, and the term thermoelectric is entered as a search concept. Thermoelectrics is an area of research that focuses on materials conductivity of temperature (heat or cooling) for energy production. In this example, the term was only found in the LCSH, which is a general domain ontology. The lower-half of Figure \ref{fig3} shows the term relationships. There are other tabs accessible to see the JSON-LD, SKOS-RDF/XML and other encoding. This feature also allows a user to select an encoded term for a structure database system, such as a catalog, or for inclusion in a knowledge graph.


\begin{figure}
\centering
\includegraphics[width=\textwidth]{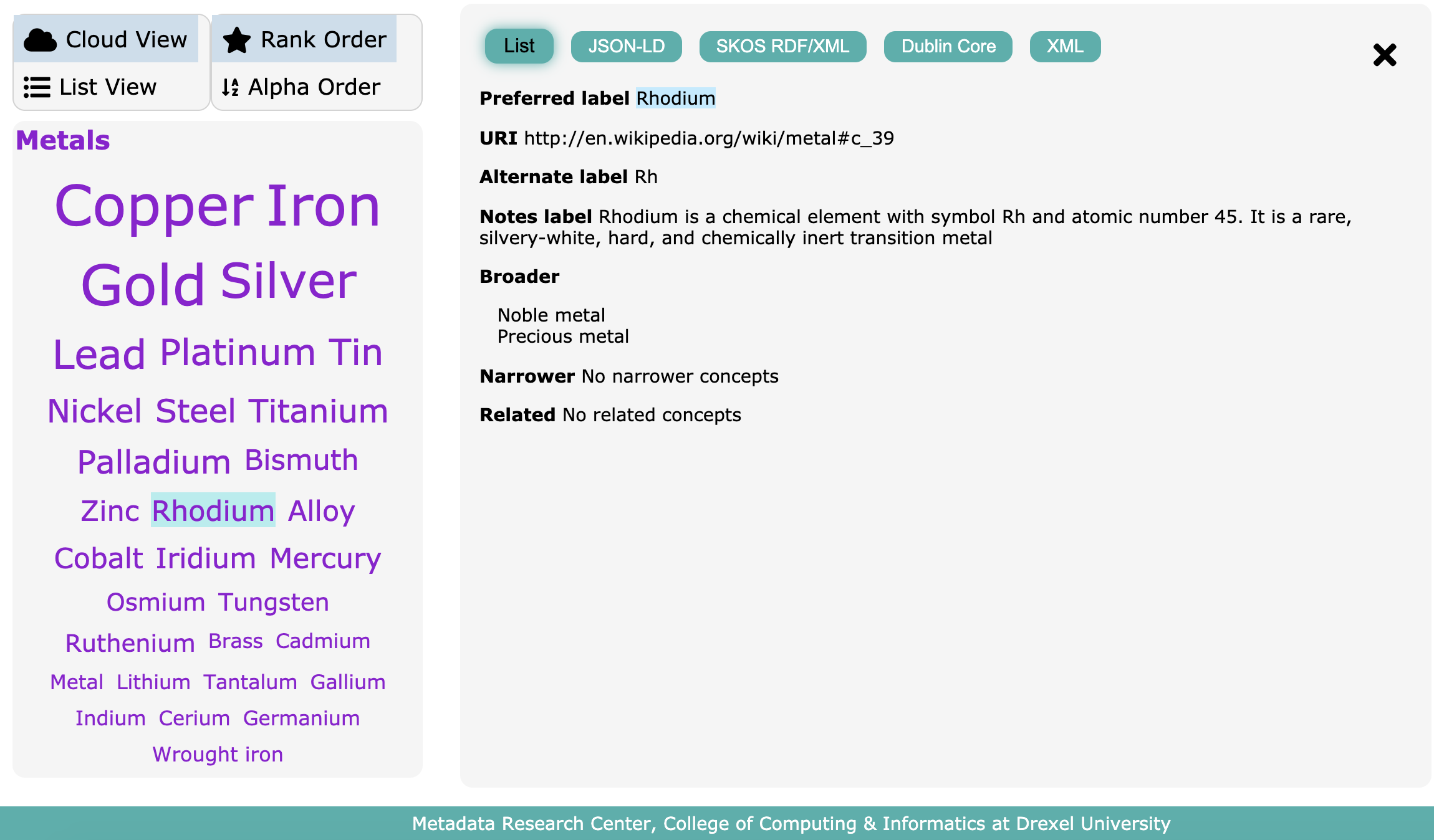}
\caption{Keyword Extraction}
\label{fig4}
\end{figure}

Figure \ref{fig4} illustrates the Knowledge Extraction/Indexing Feature. To reiterate, reading research literature is time-consuming. Moreover, it is impossible for a researcher to fully examine and synthesize all of the knowledge from existing work. HIVE-4-MAT’s indexing functionality allows a researcher or a digital content curator to upload a batch of textual resources, or simply input a uniform resource locator (URL) for a web resource, and automatically index the textual content using the selected ontologies. Figure \ref{fig4} provides an example using the Wikipedia content for Wikipedia page on Metal\cite{Metal}. The visualization of the HIVE-4-MAT’s results helps a user to gain an understanding of the knowledge contained within the resource, and they can further navigate the hypertext to confirm the meaning of a term within the larger ontological structure.

\subsection{Building NER for Information Extraction}

Inspired by the work of Weston et al.\cite{Weston2019}, the HIVE team is also exploring the performance and applications of NER as part of knowledge extraction in materials science. Research in this area may also serve to enhance HIVE. Weston et al.\cite{Weston2019} focus on inorganic materials, and appear to be one of the only advanced initiative's in this area. Our current effort focuses on building a test dataset for \textbf{organic materials} discovery, with the larger aim of expanding research across materials science. 

\begin{figure}
\centering
\includegraphics[width=\textwidth]{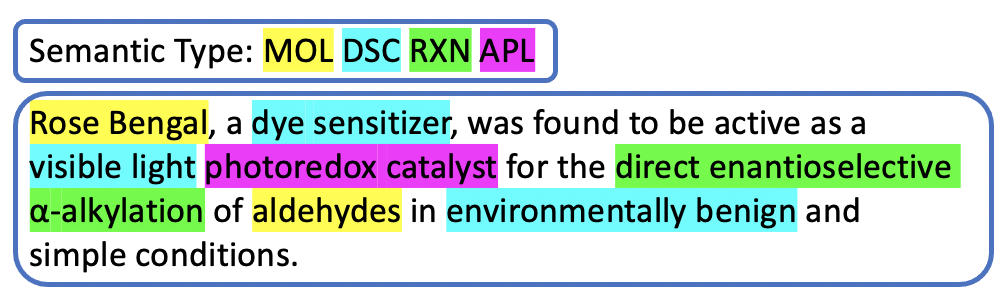}
\caption{Example from our In-Progress Organic Dataset}
\label{organic example}
\end{figure}

To build our corpus, we used Scopus API\cite{kaiwan:2019} to collect a sample of abstracts from a set of journals published by Elsevier that cover organic materials. The research team has identified and defined a set of seven key entities to assist with the next step of of training our model. These entities have the following semantic labels: (1) Molecules/fragments, (2) Polymers/organic materials, (3) Descriptors, (4) Property, (5) Application, (6) Reaction and (7) Characterization method. Members of our larger research team are actively annotating the abstracts using these semantic labels as shown in Figure \ref{organic example}. The development a test dataset is an important research step, and will help our team move forward testing our NER model and advancing knowledge extraction options for materials science in our future work.

\section{Discussion}

The demonstration of HIVE and reporting of initial work with NER is motivated by the significant challenge materials science researchers face gleaning knowledge from textual artifacts. Although this challenge pervades all areas of scientific research, disciplines such as biology, biomedicine, astronomy, and other earth sciences have a much longer history of open data and ontology development, which drives knowledge discovery. Materials science has been slow to embrace these developments, most likely due to the disciplines connection with competitive industries. Regardless of the reasons impacting timing, there is clearly increased interest and acceptance of a more open ethos across materials science, as demonstrated by initiatives outlined by Himanen et al. in 2019 \cite{himanen2019data}. Two key examples include NOMADCoE \cite{draxl2018nomad} and the Materials Data Facility \cite{blaiszik2016materials}, which are inspired by the FAIR principles \cite{draxl2018nomad,wilkinson2016fair}. These developments provide access to structured data, although, still the majority of materials science knowledge remains hidden in textually dense artifacts. More importantly, these efforts recognize the value of access to robust and disciplinary relevant ontologies. HIVE-4-MAT complements these developments and enables materials science researchers not only to gather, register, and browse ontologies; but, also the ability to automatically apply both general and targeted ontologies for knowledge extraction. Finally, the HIVE-4-MAT output provides researchers with a structured display of knowledge that was previously hidden within unstructured text.

\section{Conclusion}

This paper introduced the HIVE-4-MAT application, demonstrated HIVE's three key features, and reported on innovative work underway exploring NER. The progress has been encouraging, and plans are underway to further assess the strengths and limitation of existing ontologies for materials science. Research here will help our team target areas where richer ontological structures are needed. Another goal is to test additional algorithms with the HIVE-4-MAT application, as reported by White, et al\cite{white2012hive}. Finally, as the team moves forward, it is critical to recognize that ontologies, alone, are not sufficient for extracting knowledge, and it is important to consider other approaches for knowledge extraction, such as Named Entity Recognition (NER) and Relation Extraction (RE) can complement and enrich current apporaches. As reported above, the HIVE team is also pursuing research in this area as reported by Zhao\cite{zhao2020scholarly}, which we plan to integrate with the overall HIVE-4-MAT.

\section{Acknowledgment}

The research reported on in this paper is supported, in part, by the U.S. National Science Foundation, Office of Advanced Cyberinfrastructure (OAC): Grant: \#1940239. Thank you also to researchers in Professor Steven Lopez's lab, Northeastern University, and Semion Saiki, Kebotix for assistance in developing the entity set for organic materials.

%
%
%
%
\bibliographystyle{plain} 

\end{document}